\documentclass[a4paper,11pt]{article}
\usepackage{jcappub} 
\usepackage{siunitx}

\usepackage[super]{nth}
\usepackage{xcolor, soul}
\usepackage{orcidlink}

\newcommand\arcmin{\mbox{$^\prime$}}%
\usepackage[normalem]{ulem}

\arxivnumber{2406.11992} 
\title{Modeling optical systematics for the Taurus CMB experiment}

\author[a,b]{Alexandre E. Adler\orcidlink{0000-0002-5736-5524},}
\author[c]{Jason E. Austermann,}
\author[d]{Steven J. Benton\orcidlink{0000-0002-4214-9298},}
\author[c]{Shannon M. Duff,}
\author[e]{Jeffrey P. Filippini\orcidlink{0000-0001-8217-6832},}
\author[d]{Aurelien A. Fraisse,}
\author[f]{Thomas Gascard\orcidlink{0000-0002-7020-7977},}
\author[e]{Sho M. Gibbs\orcidlink{0009-0009-8339-5231},}
\author[d]{Suren Gourapura\orcidlink{0000-0002-8149-0632},}
\author[c]{Johannes Hubmayr\orcidlink{0000-0002-2781-9302},}
\author[f,g]{Jon E. Gudmundsson\orcidlink{0000-0003-1760-0355},}
\author[d]{William C. Jones,}
\author[h]{Jared L. May\orcidlink{0000-0003-3402-488X},}
\author[h]{Johanna M. Nagy\orcidlink{0000-0002-2036-7008},}
\author[c,i]{Kate Okun,}
\author[h]{Ivan Padilla,}
\author[c]{Christopher Rooney,}
\author[d]{Simon Tartakovsky\orcidlink{0009-0006-8752-1424},}
\author[c]{Michael R. Vissers.}
\affiliation[a]{Department of Physics, University of California, Berkeley, 366 LeConte Hall Berkeley, CA 94720, USA}
\affiliation[b]{Physics Division, Lawrence Berkeley National Laboratory, 
1 Cyclotron Road, Berkeley, CA 94720, USA}
\affiliation[c]{Quantum Sensors Division, National Institute of Standards and Technology, Boulder, CO 80305, USA}
\affiliation[d]{Department of Physics, Princeton University, Jadwin Hall, Princeton, NJ 08544, USA}
\affiliation[e]{Department of Physics, University of Illinois Urbana-Champaign, 1110 W Green St, Urbana, IL 61801, USA}
\affiliation[f]{Science Institute, University of Iceland, 107 Reykjavik, Iceland}
\affiliation[g]{The Oskar Klein Centre, Department of Physics, Stockholm University, AlbaNova, SE-10691 Stockholm, Sweden}
\affiliation[h]{Department of Physics, Case Western Reserve University, 10900 Euclid Ave, Cleveland, OH 44106, USA}
\affiliation[i]{ Department of Astrophysical and Planetary Sciences, University of Colorado, 2000 Colorado Ave, Boulder, CO 80309, USA}

\emailAdd{aadler@lbl.gov}

\abstract{We simulate a variety of optical systematics for Taurus, a balloon-borne cosmic microwave background (CMB) polarisation experiment, to assess their impact on large-scale $E$-mode polarisation measurements and constraints of the optical depth to reionisation $\tau$. 
We model a one-month flight of Taurus from Wanaka, New Zealand aboard a super-pressure balloon (SPB). 
We simulate night-time scans of both the CMB and dust foregrounds in the \SI{150}{\giga\hertz} band, one of Taurus's four observing bands.
We consider a variety of possible systematics that may affect Taurus's observations, including non-gaussian beams, pointing reconstruction error, and half-wave plate (HWP) non-idealities. 
For each of these, we evaluate the residual power in the difference between maps simulated with and without the systematic, and compare this to the expected signal level corresponding to Taurus's science goals.
Our results indicate that most of the HWP-related systematics can be mitigated to be smaller than sample variance by calibrating with Planck's $TT$ spectrum and using an achromatic HWP model, with a preference for five layers of sapphire to ensure good systematic control. 
However, additional beam characterization will be required to mitigate far-sidelobe pickup from dust on larger scales.}

\begin{document}
\maketitle
\flushbottom


\section{Introduction} \label{sec:intro}
Since the mid-1990s a concordance model of cosmology has emerged, known as “Lambda cold dark matter” ($\Lambda $CDM). 
The most basic $\Lambda$CDM model, defined with only six parameters, has been found to fit a wide range of observations from large-scale structure, supernovae and the cosmic microwave background (CMB) rather well \cite{Peebles2020}. 
Each of these parameters tells us specific information about our Universe, and can also help constrain fundamental physics.
Some of the strictest constraints on them come from observations of the CMB \cite{Planck18-6}.

One of the six parameters is $\tau$, the optical depth to reionisation. 
At face value, $\tau$ tells us the integrated density of free electrons along the line of sight since the end of reionisation, or equivalently the redshift $z_\mathrm{re}$ at which an instantaneous reionisation would have taken place. 
During and after reionisation, CMB photons can scatter on free electrons. 
In the non-relativistic limit, this reduces to Thomson scattering that suppresses temperature anisotropies in the CMB by a factor of $e^{-2\tau}$ at the power spectrum level.
However, the differential Thomson scattering cross-section is polarisation dependent. 
An electron seeing a local quadrupolar anisotropy of source photons will re-scatter them with a net linear polarisation \cite{Zaldarriaga1997, HaimanKnox1999, Aghanim2006, Aghanim2008}. 
Since our lines of sight as observers point to the last scatterings of the individual photons, the net linear polarisation from any particular line of sight will then depend on the CMB quadrupole moment as it appeared on the cosmological horizon at the location of that last scattering \cite{Planckint-48}.
The horizon size at reionisation subtends a large area on the sky, therefore we expect reionisation to induce polarisation correlations on large angles, which will lead an observer today to see a small excess in the polarisation power spectrum at small multipoles. 

Knowledge of $\tau$ also informs attempts to estimate the sum of neutrino masses from the CMB. 
Primordial density perturbations seed structure growth, which is damped by relativistic neutrinos streaming out of areas of high gravitational potential \cite{Hu1998}. 
Since massive neutrinos become non-relativistic at late times, precision measurements of this damping as a function of scale (e.g., from gravitational lensing of the CMB) can be used to constrain the sum of neutrino masses, $\Sigma m_\nu$ \cite{Abazajian2015}. 
This requires a precise estimate of $A_s$, the amplitude of the primordial density perturbations, but there is a degeneracy between $A_s$ and $\tau$, as the amplitude of the CMB temperature power spectrum is a multiple of $A_s e^{-2\tau}$ \cite{WeinbergCosmo7.2}. 
A precise estimate of $\Sigma m_\nu$ from the CMB thus relies on an estimate of $\tau$.
Today, the best constraints on $A_s$ and $\tau$ come from the \textit{Planck} satellite \cite{Planck18-6, Planckint-57, Tristram2024}. 
While $A_s$ is determined to percent level accuracy, \textit{Planck}'s constraints on $\tau$ are an order of magnitude looser, with $\sigma(\tau)=0.006$.
This motivates an experiment dedicated to constraining $\tau$ independently from \textit{Planck}.
Furthermore, in the coming decade, new observations will change our understanding of galaxies and stars in the epoch of reionisation \cite{SKA2015}. 
This new history of reionisation can be compared with the CMB inferred $\tau$ in a valuable cross-check of $\Lambda$CDM.

\begin{figure}
    \centering
    \includegraphics[width=\textwidth]{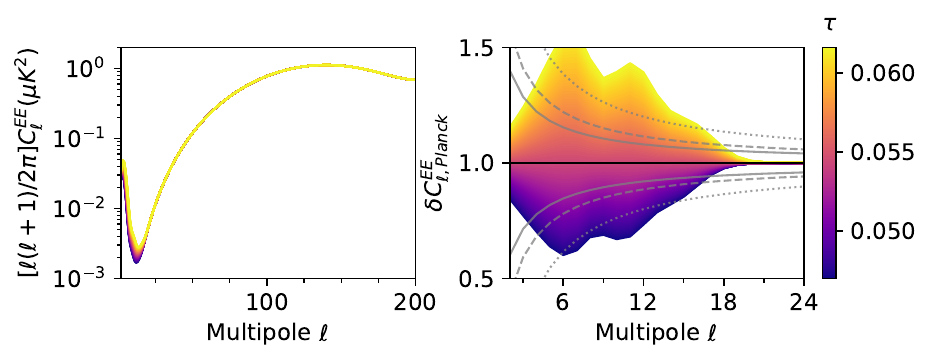}
    \caption{Left: the CMB $EE$ power spectrum for $2\leq\ell\leq200$ for a variety of values of $\tau$ within Planck's $1\sigma$ uncertainty \cite{Planck18-6} (color bar), with the product $A_s e^{-2\tau}$ kept constant. 
    Right: the relative variation from the \textit{Planck} nominal value for $2 \leq \ell \leq 24$.
    The grey lines represent the sample variance associated with the $EE$ spectrum for different observed sky fractions: the dotted line is for $f_\textrm{sky}=0.4$, the dashed line for $f_\textrm{sky}=0.7$ and the full line is the cosmic variance limit on $\sigma(\tau)$. 
    In our simulations, we estimate the power spectrum on roughly \SI{40}{\percent} of the sky.}
    \label{fig:EEspectrum}
\end{figure}

In Figure~\ref{fig:EEspectrum} we can see how the amplitude of the $EE$ spectrum at large angular scales depends on the value of $\tau$ even if $A_se^{-2\tau}$ stays constant.
\textit{Planck}'s $1\sigma$ uncertainty on $\tau$ is larger than the $EE$ spectrum's sample variance for $5<\ell<15$. 
Then an experiment that can measure the $EE$ power spectrum around $\ell\sim 10$ could  constrain $\tau$.
Beyond $\ell=20$, any variation in $D_\ell^{EE}$ due to the uncertainty on $\tau$ is compatible with cosmic variance. 
Depending on the observed sky fraction, the signal starts being dominated by sample variance for $15\leq\ell\leq20$.
Therefore, to probe the angular power spectra on large scales, an experiment with a wide sky coverage is needed.
Furthermore, to disentangle polarized foregrounds from the CMB, multi-frequency maps are required.
Studies of our Galaxy will also benefit from foreground polarisation maps at higher-than-\textit{Planck} sensitivity. 
Taurus, a balloon-borne CMB experiment flying at mid-latitudes, can assure both the wide spatial and spectral coverage.
Taurus' current configuration is described in detail in \cite{May2024}, and its cryostat design in \cite{Tartakovsky2024}.
To ensure the success of Taurus, the mission must be designed to minimize systematics.
Modeling can inform us, however incompletely, on the amount of systematic error that the current Taurus design will face on deployment, and therefore influence the evolution of that design.
In this paper, we simulate time-ordered data and sky maps for a Taurus mission profile, evaluating the impact that beam and HWP systematics might have. 
In Section~\ref{sec:taurus}, we introduce the Taurus mission.
Section~\ref{sec:model} describes how we simulate a Taurus flight. 
In Section~\ref{sec:pipeline} we describe the treatment of the simulated sky maps and introduce the various calibration parameters we apply.
Results are presented in Section~\ref{sec:results}, and we discuss their meaning for Taurus in Section~\ref{sec:conclusion}, outlining our future lines of inquiry.

\section{Taurus: Experiment design}\label{sec:taurus}
The baseline Taurus design has three cryogenically cooled receivers that will observe the sky at four frequencies using arrays of superconducting transition-edge sensor (TES) detectors coupled to feedhorns.
All the feedhorn pixels are dichroic, meaning they feed two TESs sensitive to two distinct frequencies. 
This increases the amount of information that can be gathered from a given focal plane area. 
The four bands are grouped in pairs, with the bands centred at 150 and \SI{220}{\giga\hertz} designated LF1 and LF2 respectively, while the bands centred at 280 and \SI{350}{\giga\hertz} are HF1 and HF2. 
A batch of detectors, split between two receivers, observes the LF bands.
Another batch, grouped in one focal plane, observes the HF bands.
Each of the three receivers is a refractor with a \SI{28}{\degree} field of view.
There will be a polarisation modulator in front of each telescope's stop: due to the dichroic nature of the receivers, a broadband half-wave plate (HWP) is required. 
The HWP is stepped regularly to improve cross-linking of the observations and mitigate beam systematics.
The three receivers are offset by \SI{90}{\degree} in azimuthal pointing from one another.

Taurus will fly suspended from a super-pressure balloon (SPB) launched from New Zealand.
We are planning for the flight duration to be of order one month, during which the balloon circumnavigates at mid-latitudes pushed by the prevailing stratospheric winds (as represented in Figure~\ref{fig:position}).
During the day, Taurus will recharge its batteries by pointing its solar panels towards the Sun.
Thanks to the long flight duration, limiting Taurus to night-time observations only modestly reduces the accessible sky coverage.
The instrument rotates about its axis and observes at a fixed boresight elevation of \SI{35}{\degree}; because Taurus scans at night, it does not need to avoid the Sun and can rotate through the whole \SI{360}{\degree} azimuthal range.
The payload rotates at \SI{30}{\degree\second^{-1}}.
The HWPs are stepped once per day by integer multiples of \SI{22.5}{\degree}. 
With such a scan strategy, a one-month flight will get well-conditioned maps as shown in Figure~\ref{fig:hitsmap}: in our simulations, the condition number is smaller than $2.3$ for \SI{90}{\percent} of the observed pixels and smaller than $3$ for \SI{99}{\percent} of them. 

\section{Model Parameters}\label{sec:model}

\subsection{Modeling the flight}
\begin{figure}
    \centering
    \includegraphics{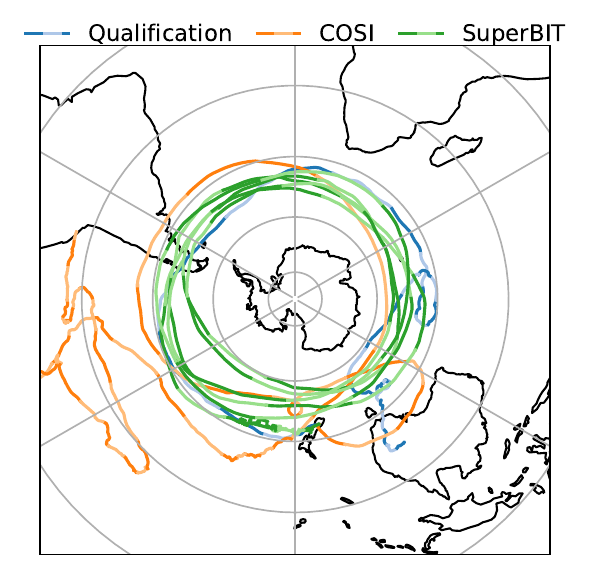}
    \caption{Trajectories for the three longest NASA super-pressure balloon flights to date from Wanaka, NZ \cite{Cathey2017, Kierans2017, Sirks2023}. Alternating shading along each track demarcates each day of flight. Our simulated trajectory follows the blue track (qualification) for thirty days.}
    \label{fig:position}
\end{figure}

\begin{figure}
    \centering
    \includegraphics[width=\textwidth]{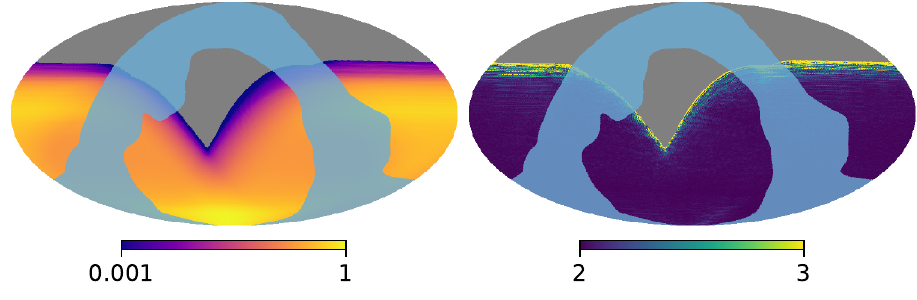}
    \caption{Left: Normalized hits map, in equatorial coordinates, for the scanning strategy described in Section~\ref{sec:taurus} for 49 detector pairs sampling at \SI{50.01}{\hertz}. Grey pixels are not observed.
    The cyan area falls within our galactic mask and covers the \SI{30}{\percent} of the sky most contaminated by the Galaxy according to \textit{Planck}.
    Right: Condition number map.}
    \label{fig:hitsmap}
\end{figure}

There have been five SPB launches from Wanaka in New Zealand since the Columbia Scientific Ballooning Facility began its operations there in 2015. 
They are designed to minimize altitude fluctuations due to day/night temperature cycles. 
The SPBs had long hold times for three out of five flights: the qualification flight lasted 32 days \cite{Cathey2017}, the SuperBIT flight was 39 days long \cite{Sirks2023}, and COSI stayed aloft for 46 days \cite{Kierans2017}. 
Their trajectories are represented in Figure~\ref{fig:position}.
For the purposes of this work, we have chosen to assume the trajectory of Flight 662 NT, the qualification flight, which had an average altitude of \SI{33500}{\metre}. 
Our simulation covers the first 30 days of that flight, from March $26^\textrm{th}$ to April $26^\textrm{th}$. 
We let the simulated payload's altitude vary daily around an average value of \SI{32700}{\metre}, with a standard deviation of \SI{200}{\metre}.
The resulting sky coverage is visible in the left panel of Figure~\ref{fig:hitsmap}.

At \SI{45}{\degree}S, roughly \SI{85}{\percent} of the sky is visible over a 24-hour period. 
Not all of this is visible for a balloon-borne telescope, as the atmosphere degrades performance at low elevation, and the balloon obscures the zenith. 
With Taurus' pointing and field of view, the maximum fraction of the sky visible is $f_\mathrm{sky}\approx\SI{80}{\percent}$.
Depending on the season of the launch, parts of the sky will be only visible during the day, reducing the observed sky fraction.
We follow the civil definitions of sunset/sunrise (center of the solar disk \SI{5}{\degree} below the horizon), which we compute for a given position and altitude along the trajectory using PyEphem \cite{PyEphem}.
A flight starting March $1^\textrm{st}$ will see $f_\mathrm{sky} = \SI{62}{\percent}$ while a June launch would get $f_\mathrm{sky} = \SI{70}{\percent}$. 
The input CMB maps are generated from the \textit{Planck} power spectrum with \texttt{healpy}\footnote{http://healpix.sourceforge.net} \cite{Gorski2005, Zonca2019}, while the dust emission is based on the \texttt{PySM} \cite{Thorne2017} \texttt{d9} model.
In that model, the dust emission is given by a modified blackbody whose temperature $T_d$ and spectral index $\beta$ vary from pixel to pixel.
While more complex models exist in \texttt{PySM}, it is a good starting point to assess the impact of differences between the CMB and the dust on our instrument's performance.

\subsection{Focal plane modeling}
We model beams for 49 detector pairs of one of the LF receivers, covering a \SI{20}{\degree} $\times$ \SI{20}{\degree} square grid on the sky. 
The diagonal therefore covers the instrument's \SI{28}{\degree} field of view.
Each detector pair has two detectors that are aligned with $\pm Q$ or $\pm U$ (i.e. the detectors' polarisation angles are \SI{90}{\degree} apart).
Neighboring detector pairs are staggered by \SI{45}{\degree} to alternate between $Q$ and $U$ sensitivity.
The Taurus optical design is still evolving, but the baseline refractor designs gives the beams a FWHM of at least \SI{22.5}{\arcmin} (for the \SI{220}{\giga\hertz} band) and at most \SI{33}{\arcmin} (for the \SI{150}{\giga\hertz} band). 
Values for all four frequency bands are given in Table~\ref{tab:beam_FWHM}.
In this study, we focus on simulations of the \SI{150}{\giga\hertz} band.

For added realism, we use the TICRA-tools GRASP package\footnote{https://www.ticra.com/ticratools/} to simulate Physical Optics (PO) beam models for each of the detector positions we study.
The telescope model in GRASP comprises full-wave simulations of detector horns, two HDPE lenses, and a circular aperture representing the vacuum window.
The aperture is \SI{228}{\milli\metre} in diameter with an $f$-number of $f/1.6$, and the lenses focus the light down to a $\SI{200}{\milli\metre}\times\SI{200}{\milli\metre}$ square focal plane.
The fast optics result in a \SI{-10}{\deci\bel} edge taper averaged across the field of view.
We neglect reflections on the walls of the optics tube.
A Lambertian source at the aperture emits \SI{-20}{\deci\bel} of unpolarized power relative to the power passing through the aperture due to the input feedhorn beam. 
This is done to simulate scattering by the vacuum window or other optical elements near the aperture.
The aperture also acts as the overall stop of the system.
GRASP outputs both co- and cross-polar far-field beams, which can then be translated to Stokes $(I,Q,U)$ beams. 
We sample the beam finely within \SI{3}{\degree} of the beam centroid (the main beam), and more loosely up to \SI{30}{\degree} (to model beam sidelobes). 
The coarser resolution on sidelobes is reasonable as they are both much weaker and extend much further than the main beam.

\begin{figure}
    \centering
    \includegraphics[width=\textwidth]{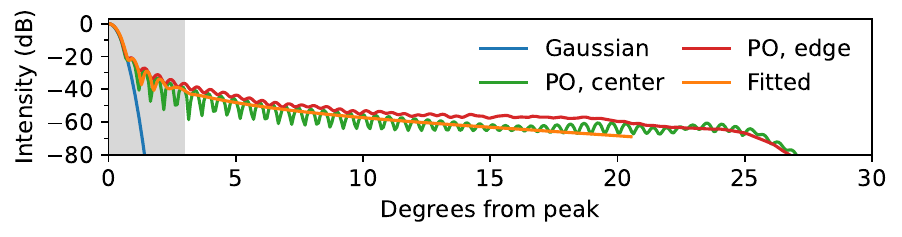}
    \caption{Azimuthally-averaged Stokes $I$ beam intensity for pixels of our simulated refractor. 
    In blue, a \SI{33}{\arcmin} FWHM Gaussian beam. 
    The green and red lines are the azimuthally averaged profiles for the PO simulations. 
    In orange, an analytical fitting of the PO beam for the central detector. 
    To evaluate the relative contribution of main lobe and side lobe for the PO beams, we can run simulations that ignore any beam power beyond \SI{3}{\degree} of the beam centroid. 
    This is represented by the gray outline.
    \label{fig:beamprofiles}}
\end{figure}

Finally, those PO beams can also be approximated by a fit model of their co-polar component. 
We adopt the fitting algorithm described in \cite{Lungu2022} and \cite{Dachlythra2024} to create a symmetrized beam model of the central pixel's PO beam.
An azimuthally averaged beam profile for the different beam models can be seen in Figure~\ref{fig:beamprofiles}.

\subsection{Half-wave plate}

A half-wave plate modulates polarisation by introducing a phase of $\pi$ radians between the two linear polarisations.
We often describe the incoming radiation using the transposed Stokes vector $S^\mathrm{t}=(I, Q, U, V)$, where $I$ is total intensity, $Q, U$ are the two linear polarisations, and $V$ is circular polarisation. 
These four Stokes parameters are real-valued, with $I\geq \sqrt{Q^2+U^2+V^2}$. 
The action of the HWP on the Stokes vector is expressed by means of a Mueller matrix $M_{HWP}$, a $4\times4$ square matrix that transforms $S$ \cite{Gil2016, Hecht2002}.
For an ideal HWP, the Mueller matrix is diagonal, and 
\begin{equation}
M_{HWP}S = 
\begin{pmatrix} 1 & 0 & 0 & 0\\
                 0 & 1 & 0 & 0\\
                 0 & 0 & -1& 0\\
                 0 & 0 & 0 &-1\end{pmatrix}
\begin{pmatrix}
    I \\ Q \\ U \\ V
\end{pmatrix} = 
\begin{pmatrix}
    I \\ Q \\ -U \\ -V
\end{pmatrix}.
\end{equation}
When the ideal HWP rotates by an angle $\theta$, the Stokes vector then gets modulated: 
\begin{equation}\label{eq:hwp_ideal_rot}
    M_{HWP}(\theta) S = \begin{pmatrix}
        I \\ Q\cos(4\theta)+U\sin(4\theta) \\ -Q\sin(4\theta)-U\cos(4\theta) \\ -V
    \end{pmatrix} 
\end{equation}
A real HWP will have a more complex Mueller matrix: the diagonal elements will no longer be $\pm 1$, and the other elements can be non-zero \cite{Bryan2010}. 
The off-diagonal elements will cause temperature to polarisation leakage, or complicate the mixing between the $Q$ and $U$ terms in Equation~\ref{eq:hwp_ideal_rot} \cite{Kusaka2014, Hill2016}. 
Furthermore, the elements of the Mueller matrix will be frequency dependent, so those non-idealities will be different for every band we observe, and can even be tricky to model within a band. 
It is therefore in our interest to design a HWP which has a stable behavior over a wide frequency band: an achromatic HWP (AHWP) \cite{Pancharatnam1955}.

Each of the Taurus HWPs is designed to be achromatic, in order to cover the two LF or HF bands. 
AHWPs can be created either by stacking layers of birefringent materials with their ordinary axes at specific angles \cite{Pancharatnam1955, Title1975}, or by using a wire mesh acting as a meta-material \cite{Pisano2016}.
For Taurus, our simulations assume one, three or five layers of birefringent sapphire, which we will refer to as BR1, BR3, and BR5 respectively. 
Each HWP is covered in three layers of dielectric anti-reflection coating\footnote{It is also possible to manufacture ablated meta-material AR coatings \cite{Takaku2020}.}, whose dielectric constant gradually increases from that of free-space to that of sapphire $(n_1, \: n_2, \: n_3)=(1.27, \: 1.98, \: 2.86)$. 
The corresponding thicknesses of the AR coating layers are 0.34, 0.21, and 0.18 \SI{}{\milli\metre}.
BR3 is based on the POLARBEAR-2 HWP \cite{Hill2016}, BR5 on a potential HWP for LiteBIRD \cite{Komatsu2020}. 
At a fixed frequency, $\nu$, the optimal thickness, $d$, of a sapphire layer is given by:
\begin{equation}
    d = \frac{c}{2\Delta n \nu} \: ,
\end{equation}
with $c$ the speed of light in a vacuum and $\Delta n$ the difference between the two indices of refraction of sapphire. 
In our simulations, $\nu$ is the mid-point between the two bands of LF or HF.
All the sapphire layers in our HWP models have the same thickness for a given receiver: the values are given in Table~\ref{tab:beam_FWHM}.
We calculate the Mueller matrices for stacks of dielectric and birefringent materials using software described in \cite{EssingerHileman2013}. 

\begin{table}
    \centering
    \begin{tabular}{|c|cccc|}\hline
         Receiver & \multicolumn{2}{c}{LF} & \multicolumn{2}{c|}{HF} \\\hline
         Band center (\SI{}{\giga\hertz}) & 150 & 220 & 280 & 350  \\\hline 
         Beam FWHM  (\SI{}{\arcmin}) & 33 & 22.5 & 30 & 25\\\hline
         Sapphire layer thickness (\SI{}{\milli\metre}) & \multicolumn{2}{c}{2.55} & \multicolumn{2}{c|}{1.50} \\ \hline
    \end{tabular}
    \caption{Frequency-dependent properties of the instrument model.}
    \label{tab:beam_FWHM}
\end{table}

\subsection{The beamconv library}\label{ssec:beamconv}
Our modelling is built around \textit{beamconv} \cite{Duivenvoorden2018}, a lightweight algorithm/python library to simulate time-ordered data from beam-convolved CMB maps. 
In \textit{beamconv}, both the beam and the sky map are decomposed into spin-weighted spherical harmonics.
The coefficients can be linearly combined into spin maps, which are then sampled by \textit{beamconv} following a sequence of pointings that represent the telescope's scan strategy, appropriately modulated by the HWP Mueller matrix.
We therefore obtain time-ordered data (TOD) for each detector that gets projected into a map using a simple binning scheme.
A real flight will discard a moderate fraction of this TOD due to fridge cycles, antenna noise, and other unavoidable instrumental noise, but ignoring this won’t bias our simulations.
To simulate the variation in either beam or sky behavior over a band, \textit{beamconv} simulations for several sub-frequencies within a band can be co-added into a band-averaged map.
When testing for frequency-dependent systematics, like HWP non-idealities or the influence of dust, we make nine simulations spaced every \SI{5}{\giga\hertz} between 130 and \SI{170}{\giga\hertz}. 
For each frequency, the behavior of the HWP and the sky will be different: we do not simulate variations of the beam within the band. 
\textit{beamconv} has been used in the recent past to estimate the effect of HWP non-idealities on $B$-mode and cosmic birefringence searches \cite{Duivenvoorden2021, Monelli2023}.

\section{Analysis of output maps\label{sec:pipeline}}
\subsection{Power spectrum estimation}

Power spectrum estimation is fundamentally limited by cosmic variance \cite{Kamionkowski1997, Louis2019}.
Beyond cosmic variance, power spectrum estimators acting on cut-sky data are sensitive to several effects due to degeneracies among spherical harmonics when measured on only part of the sky. 
To choose a specific estimator, we quantify the effects of mode-mixing and $EB$ leakage. 
We simulate 100 realisations of the CMB based on Planck's best fit power spectrum \cite{Planck18-5} and then mask areas outside of Taurus' scan range (see Figure~\ref{fig:hitsmap}). 
We also mask the \SI{30}{\percent} of the sky most contaminated by the galactic plane, resulting in around \SI{44}{\percent} of the sky remaining unmasked.
We estimated the power spectrum of the masked maps, downgraded to a resolution of NSIDE=8, using three estimators: Polspice \cite{Chon2004}, xQML \cite{Vanneste2018}, and Namaster \cite{Alonso2019}. 
We decide to use Namaster (without $B$-mode purification) for the rest of our analysis, as it recovers the input $EE$ spectrum with relatively little bias over the multipole range $2\leq \ell \leq 22$.
xQML, while more consistent in its estimations, exhibits a bias.
We note that after mode-coupling is corrected, the auto-spectra in Namaster are no longer necessarily positive on the very large angular scales.
Since the reionisation bump in the $EE$ spectrum is fairly narrow, we decide to divide the spectrum into bins of width $\Delta \ell = 5$. 
The reionisation signal will be contained within the first three bins, while other bins can be used for calibration.

To quantify the effect of non-idealities, we will look at the differences between maps simulated with a systematic (non-ideal maps) and other maps simulated without it (ideal maps). 
By estimating the power spectrum of the difference maps we can gauge the effect of the systematic at the $C_\ell$ level. 
We can then compare that effect to the uncertainty in the $EE$ power spectrum that we expect from sample variance for the observed sky fraction: $C_\ell^{EE} /\left[(2\ell+1) f_{\mathrm{sky}}\right]$.
We build our sample variance estimate following the Planck best-fit value $\tau=0.0543$ \cite{Planck18-6} and $f_\textrm{sky}=0.44$. 

\subsection{Half-wave plate efficiency} \label{ssec:hwpeffcal}
All microwave telescopes rely on some form of instrument calibration to make sense of the data they collect. 
In particular, they need to calibrate the beam of the instrument and the efficiency $\epsilon$ with which the instrument turns incoming radiation into measured Stokes $I$, $Q$, and $U$.
This process involves the use of well-understood sources such as planets \cite{Dachlythra2024}, thermal sources on drones and balloons \cite{Coppi2022}, the CMB dipole \cite{Planck15-8}, or lab-based holography \cite{Chesmore2022} to provide estimates of the efficiency and construct beam maps.
In the presence of a polarisation modulator like a HWP, the efficiency will depend on the non-idealities of that modulator, as represented by its Mueller matrix.
The efficiency will be different for total intensity and  polarisation: we can define an $\epsilon_I$ for intensity and an $\epsilon_P$ for the polarisation modulation efficiency, with $|P|=\sqrt{Q^2+U^2}$. 

\begin{figure}
    \centering
    \includegraphics[width=\textwidth]{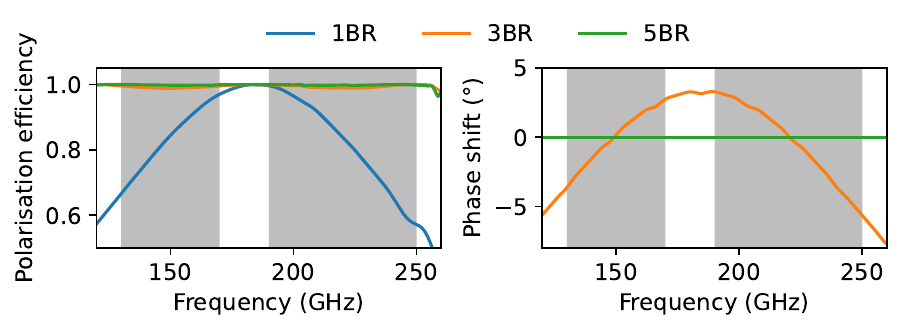}
    \caption{
    Figures of merit as a function of frequency from the Mueller matrices of the three HWP models studied, computed using \cite{EssingerHileman2013}. 
    1BR, 3BR, 5BR refer to the number of \textbf{b}i\textbf{r}efringent layers in the HWP model.
    The shaded regions are the two LF bands of Taurus.
    Left: Polarisation modulation efficiency.
    Right: frequency-dependent phase shift. We have offset the 3BR phase by \SI{32.67}{\degree} for ease of reading. The phase-shift is zero for 1BR and negligible for 5BR.}
    \label{fig:hwp_fom}
\end{figure}

In Figure~\ref{fig:hwp_fom} we plot $\epsilon_P$ for the three HWP models that we study following the formula given in \cite{Komatsu2020}:
\begin{equation}
    \epsilon_P = \frac{\sqrt{\left(M_{QQ}-M_{UU}\right)^2 +\left(M_{UQ}+M_{QU}\right)^2 }}{2M_{II}+M_{QQ}+M_{UU}} \: , \label{eq:komatsu_hwp}
\end{equation}
where the terms in the above equation all correspond to a frequency-dependent Mueller matrix element. Equation~\ref{eq:komatsu_hwp} describes the fraction of the incoming polarized light that will be modulated by $4\theta$, like in the idealized case of Eq.~\ref{eq:hwp_ideal_rot}. 
The $\epsilon_P$ of the one-layer HWP varies within each band and can be up to \SI{40}{\percent} lower than for the three or five-layer AHWP.

The right panel of Figure~\ref{fig:hwp_fom} showcases a peculiarity of AHWPs.
While it will have a near-ideal $\epsilon_P$ over a large frequency domain, an AHWP will exhibit a frequency-dependent phase shift $\delta\alpha(\nu)$, meaning that the plane of polarisation of light exiting the HWP has been rotated, changing some of the $Q$ into $U$.
This can be accounted for during map-making: we correct the HWP angle by that phase shift \cite{Moncelsi2014, Giardiello2022}.
Since it varies over the \SI{150}{\giga\hertz} frequency band, the correction operation is accurate for the band-averaged value $\overline{\delta\alpha}$ of $\delta\alpha(\nu)$ 
\begin{equation}
    \overline{\delta\alpha} = \frac{ \int_{\nu_0 - \Delta\nu /2}^{\nu_0 + \Delta\nu /2} \delta\alpha(\nu) g(\nu) I(\nu) \mathrm{d}\nu }
    {\int_{\nu_0 - \Delta\nu /2}^{\nu_0 + \Delta\nu /2} g(\nu) I(\nu) \mathrm{d}\nu },
\end{equation}
where $g(\nu)$ is the spectral response of the telescope and detector, and $I(\nu)$ the intensity of the sky signal. 
In the above expression, we assume a spectral response function that defines a bandwidth $\Delta\nu$ centered on $\nu_0$.
The value of $\delta\alpha(\nu)$ and $g(\nu)$ can be measured during pre-flight calibration.
During observations, however, the sky signal, $I(\nu)$, represents the location-dependent sum of different components (including synchrotron emission, dust, CMB) in proportions that depend on the sky position. 
This means that the best-suited value of $\overline{\delta\alpha}$ will depend on our assumptions about the spectral energy density of the sky we are observing. 
Taking for instance the BR3 HWP model, $\overline{\delta\alpha}$ is \SI{32.67}{\degree} for a pure CMB sky but \SI{32.94}{\degree} for the dust emission described by \texttt{PySM}'s d9 model. 

\subsection{Corrections at the power-spectrum level}\label{ssec:gaincal}

The multiplicative gain due to $\epsilon_I$ and $\epsilon_P$ will propagate to maps and power spectra \cite{Monelli2024}.
Ignoring the efficiency issue leads to smaller inferred amplitudes for the CMB anisotropies which obviously impacts attempts to measure $\tau$ or the other cosmological parameters.
We therefore try to account for it by defining a HWP-related gain $G^{XY}$:
\begin{equation}
    G^{XY} = \left( \frac{
    \sum_{\ell = \ell_1}^{\ell_2} \tilde{C}_\ell^{XY; \mathrm{ideal}}}
    {\sum_{\ell = \ell_1}^{\ell_2} \tilde{C}_\ell^{XY; \mathrm{non-ideal}}}\right)^{1/2}, \label{eq:hwp_gain}
\end{equation}
where $G^{XY}$ is the square root of the relative amplitude of the cross- or auto-spectrum $XY$ of simulated maps due to the presence of the HWP non-idealities, and $(\ell_1, \ell_2)$ is the calibration range. 
Multiplying the simulated maps made with a non-ideal HWP by $G^{XY}$ compensates for the HWP efficiency losses. 
$G^{TT}$ is similar to $\epsilon_I^{-2}$, while $\epsilon_P$ can be corrected by a good estimate of $G^{EE}$ and/or $G^{BB}$. 
The difference maps $\textbf{m}^\textrm{diff}$ are then given by:
\begin{equation}
    \textbf{m}^\textrm{diff} = G^{XY}\textbf{m}^\textrm{non-ideal} -  \textbf{m}^\textrm{ideal} \: .
\end{equation}

A real experiment will not know $\tilde{C}_\ell^{XY; \mathrm{ideal}}$, but it would be reasonable to construct this quantity by convolving a known $C_\ell^{XY}$ from another experiment's spectra or maps with the effective beam transfer function $B_\ell$ for the experiment:
\begin{equation}
    \tilde{C}_\ell^{XY; \mathrm{ideal}} = C_\ell^{XY; \mathrm{known}} B_\ell^2 \: .
\end{equation}

If we rely on another experiment, the calibration range has to be at multipoles where both Taurus and the other experiment can measure the $TT$ or $EE$ spectrum accurately. 
Beyond multipoles of $\ell \sim 400$, Taurus' angular resolution limits its sensitivity. 
At the other end, setting the calibration range to be the same as the multipoles we are trying to characterize ($\ell<22$) can intertwine our analysis with the systematics of previous experiments in a very direct way. 
It is difficult to make an independent measure of the $EE$ spectrum on large angular scales if our calibration efforts depend on the measured \textit{Planck} $EE$ spectrum on those scales.
In this paper, we settle on $\ell_1=50$ and $\ell_2=100$, where \textit{Planck} was able to measure the polarisation power spectrum with high accuracy \citep{Planck18-5}. 

\begin{figure}
    \centering
    \includegraphics[width=\textwidth]{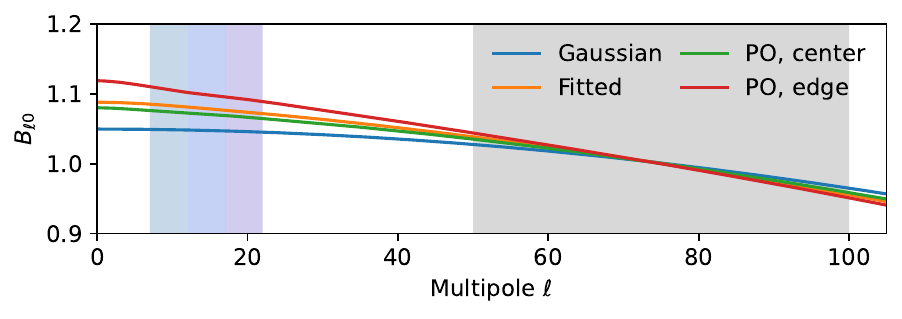}
    \caption{The azimuthal, $m=0$, component of the beam transfer functions $B_\ell$ used in our simulations, normalized to their average in the grey area, our calibration range. We can see how the $B_\ell$ diverge from it in the colored areas from $\ell=7$ to $\ell=22$ that correspond to our science bins.}
    \label{fig:bl_profiles}
\end{figure}

If the beam transfer function is correct, Eq.~\ref{eq:hwp_gain} remains unchanged. 
However, if the wrong $B_\ell$ is assumed, then $\tilde{C}_\ell^{XY; \mathrm{ideal}}$ will have an $\ell$-dependent bias that will propagate to $G^{XY}$. 
This could happen, for instance, if the assumed $B_\ell$ for the experiment ignores the beam sidelobes.
Then their $\tilde{C}_\ell$s will have a different gain in the “science” bins at  than in the calibration range, and our procedure to estimate the gain will introduce an error in the experiment's estimation of the CMB power spectrum.
Let us try to illustrate this with a simple case: suppose our ideal model assumes a Gaussian beam, denoted with g, with FWHM $f_1$. 
The transfer function for that beam is:
\begin{equation}
    B_\ell^\textrm{g} = e^{-\frac{\ell(\ell+1)f_1^2}{16\ln 2}}\label{eq:gauss_bl} \: .
\end{equation}
Suppose, however, that the true beam had an extra Gaussian sidelobe aligned with the main beam, denoted with sl, with amplitude $\varepsilon$ and FWHM $f_2$. 
This can be seen as a reasonable approximation of the "tails" of a real beam beyond the main lobe.
The transfer function of the true beam is: 
\begin{equation}
B_\ell^\mathrm{g+sl} = (1-\varepsilon)B_\ell^\textrm{g} + \varepsilon B_\ell^\textrm{sl} 
= B_\ell^\textrm{g} + \varepsilon\left( B_\ell^\textrm{sl} - B_\ell^\textrm{g}\right) \: ,
\end{equation}
where $B_\ell^\textrm{sl}$ follows the form set in Eq.~\ref{eq:gauss_bl} but with the FWHM replaced by $f_2$. 
Then, the bias per multipole is: 
\begin{equation}
    \begin{split}
    \delta \hat{C}_\ell = \frac{\hat{C}_\ell^\mathrm{g+sl} - \hat{C}^\mathrm{g}_\ell}{\hat{C}^\mathrm{g}_\ell}
    = \frac{\tilde{C}_\ell (B_\ell^{\mathrm{g+sl}})^{-2} 
    - \tilde{C}_\ell (B^\mathrm{g}_\ell)^{-2}}
    {\tilde{C}_\ell (B^\mathrm{g}_\ell)^{-2}}\\
   = \left[\frac{B^\mathrm{g}_\ell}{B_\ell^\textrm{g} + \varepsilon\left( B_\ell^\textrm{sl} - B_\ell^\textrm{g}\right) }\right]^2 -1 \approx 2\varepsilon \left(1 - \frac{B_\ell^\mathrm{sl}}{B_\ell^\mathrm{ g}}\right) + \mathcal{O}\left(\varepsilon^2\right) \: .
    \end{split}  
\end{equation}
The transfer functions for our fitted and PO beams, visible in Figure~\ref{fig:bl_profiles}, can be approximated with $\varepsilon=0.03$, $f_1=\SI{33}{\arcmin}$ and $f_2=\SI{5}{\degree}$. 
The average bias over the calibration range ($50 \leq \ell \leq 100$) is approximately \SI{-1.5e-3}{}.
In our “science” multipole range, the bias in a $7\leq \ell < 12$ multipole bin for those two beams would however be \SI{-4.4e-2}{}.
Therefore, beam mismatch during gain calibration could lead to more than \SI{4}{\percent} multiplicative bias in our estimate of the power spectrum. 

A power spectrum prior can be used for more than just efficiency calibration.
It should also be possible to calibrate the polarisation angles of the detectors by attempting to null the CMB $EB$ power spectrum \cite{Keating2013}, which is supposed to be zero in $\Lambda$CDM. 
In the same vein, \cite{Monelli2023} discussed the degeneracy between non-zero $C_\ell^{EB}$ and an incorrect estimation of the HWP angle.
We could therefore imagine determining $\overline{\delta\alpha}$ by minimizing the $EB$ power.
That is particularly attractive as polarized point sources that can be used for calibration are very rare in the microwave sky.
There are two downsides to that approach: first, it blinds us to detecting the possible non-zero CMB $EB$ correlation caused by cosmic birefringence \cite{Minami2020, Eskilt2023}.
Furthermore, the $EB$ power spectrum could contain a non-zero contribution from Galactic dust emission \cite{Clark2021}, so $EB$ minimization would not be a good option for the other bands of Taurus, which are dominated by dust emission.

\section{Results}\label{sec:results}

There is no shortage of potential instrument non-idealities that can complicate data analysis. 
With the model described in Section~\ref{sec:model}, we focus on beam and HWP non-idealities.
Beam non-idealities include deviations from the assumed beam model, errors in pointing, and polarisation angle errors. 
HWP non-idealities cover both the effect of non-ideal elements in the Mueller matrix and the coupling between those elements and the beam. 
We present the power spectrum level residuals in the multipole bin $ 7\leq \ell < 12$, comparing them to value of the sample variance for $C_\ell^{EE}$ in that bin for $f_\textrm{sky}=0.44$, which is about \SI{4.2e-5}{\micro\kelvin^2}. 

Dust emission has a different SED than the CMB. 
As discussed in Section~\ref{ssec:hwpeffcal}, this means that the optimal AHWP $\overline{\delta \alpha}$ in a given frequency band will vary depending on the observed component. 
Furthermore, the dust is not distributed isotropically on the sky. 
Most of it is in or near the Galactic plane, which we mask before taking the power spectra.
However, sidelobes could lead to emission from within the mask to leak into our observations. 
Therefore we perform the simulations of beam non-idealities, reflection ghosts, and non-ideal HWPs for the \texttt{PySM d9} sky model. 

\subsection{Optical ghosts}

Before it reaches a detector, a photon passing through a refracting telescope will interact with several optical elements, including the vacuum window of the cryostat, optical filters, the HWP, and the lenses.
Each change of medium will lead to some reflection. This can lead to optical ghosts, weaker images of the main beam which in some cases are mirrored across the boresight. 
For a beam with pointing $(x,y)$ with respect to the telescope's boresight, the associated ghost beam's centroid will be $(-x, -y)$.
While we sandwich lenses and HWP with anti-reflection coatings (ARCs) to suppress those reflections, it is challenging to totally avoid ghosting.
With a wide field of view like ours, ghosting will introduce spurious correlations at large angular scales for a single detector.
To estimate the effect of ghosting, we create simulations where each beam has a \SI{1}{\percent} reflection ghost of the main beam. 
Because of our symmetric focal plane, the ghosts are symmetrized too.
In a real experiment, dead detectors would break that symmetry and the ghosting would have a larger impact: therefore, for that part of the simulation work, we kill 7 randomly selected detector pairs out of our 49 (a roughly \SI{14}{\percent} fraction).
In our CMB simulations, the ghost beams have a negligible impact (at most a few \SI{e-4}{} of the sample variance target, \SI{4.2e-5}{\micro\kelvin^2}) for all beam models or gain calibration options.
We therefore will not discuss those simulations further. 
In Table~\ref{tab:ghost_dust} we show the impact of the ghost beams for dust scans is about \SI{6}{\percent} of our self-imposed sample variance target but is mostly unaffected by the choice of gain parameter: we only notice that the fully co-polar beam models (Gaussian, fitted) gets slightly increased residuals when calibrating on $EE$, suggesting the pickup of the additional galactic dust by the sidelobes of the ghost beams can be suppressed by calibration at intermediate scales.

\begin{table}[h]
\centering
\begin{tabular}{|c|c|c|c|} \hline 
Beam model & No cal. & $TT$ cal & $EE$ cal\\\hline
Gaussian & 0.063 & 0.060 & 0.071\\\hline
Fitted & 0.063 & 0.060 & 0.071\\\hline
PO & 0.059 & 0.057 & 0.060\\\hline
PO+Side & 0.059 & 0.056 & 0.059\\\hline
\end{tabular}
\caption{Ratio of power spectrum residuals to sample variance due to ghost beams for ideal HWP and PySM \texttt{d9} sky model, $7\leq \ell <12$ as a fraction of the sample variance in that bin.}
\label{tab:ghost_dust}
\end{table}

\subsection{Pointing and polarisation angle errors}\label{ssec:point}
Small to intermediate-level pointing reconstruction errors, also known as beam centroid errors, are known to primarily impact power spectrum reconstruction on intermediate angular scales $(500 \leq \ell\leq 1000)$ \cite{Poutanen2006}.
However, a recent reassessment in \cite{Naess2022} did show that sub-pixel errors have an impact on the $TT$ spectrum at scales $ \ell\leq 500$. 
Polarisation angle errors, both random and systematic, have been studied extensively in \cite{Salatino2011}. 
We group them with pointing errors as a polarisation-sensitive detector's pointing is defined by three angles on the sphere: $(\theta, \phi, \psi)$, where $\theta$ and $\phi$ are the two normal spherical coordinates and $\psi$ indicates the angle between the detector's orientation and the local meridian.
Alternatively, we can express these angles relative to the orientation of the telescope's boresight: $(\theta, \phi, \psi) \rightarrow (\Delta \textrm{az}, \Delta \textrm{el}, \xi)$.
The topic of polarisation angle errors has been intensely discussed recently due to the interest in cosmic birefringence \cite{Minami2020, DelaHoz2022, Eskilt2023}.

We simulate two modes of pointing error: either a common offset by \SI{1}{\arcmin} in azimuth and elevation for all detectors analogous to a bias in the boresight position or a random error for each detector drawn from a Gaussian distribution with \SI{1}{\arcmin} standard deviation. In the same vein, we examine the impact of improperly characterizing the polarisation angle of the detectors during the flight.
Here, the characteristic error is estimated to be as large as \SI{1}{\degree}. 

We find that both effects are much smaller than the sample variance in the $7\leq \ell < 12$ bin, with the random az-el offset being at most \SI{2}{\percent} of the sample variance if the simulation includes PO beams with sidelobes and we attempt to calibrate on the $EE$ spectrum.
This is consistent with the simulations done for other balloon CMB experiments, like SPIDER \cite{MacTavish2008, Odea2011}, and LSPE \cite{LSPE2021}.
The polarisation angle offset also returns very small residuals.
The negative sign is due to intrinsic variance from the Namaster mode decoupling algorithm, but the “true” value should be around the same amplitude. 

\subsection{Beam non-idealities}\label{ssec:beam_mis}

Perturbations to the Gaussian beam model are to be expected in realistic instrument designs. 
The main causes are non-uniform illumination of the aperture and extra reflections within the instrument. 
Ray-tracing or physical optics simulations can help us understand those extra components.
While they are much weaker than the main beam, the large angular extension of sidelobes will cause a detector to pick up some stray radiation from other much brighter sources, like the Galaxy or the ground.
Our physical optics (PO) simulations of the Taurus optics show an extended sidelobe with an amplitude between \SI{-40}{\deci\bel} and \SI{-60}{\deci\bel}, visible in Figure~\ref{fig:beamprofiles}.
The azimuthally averaged beam main beam profile is similar to a Gaussian beam with a \SI{33}{\arcmin} FWHM, as can be seen in Figure~\ref{fig:bl_profiles}.

\begin{figure}
\centering
\includegraphics[width=\textwidth]{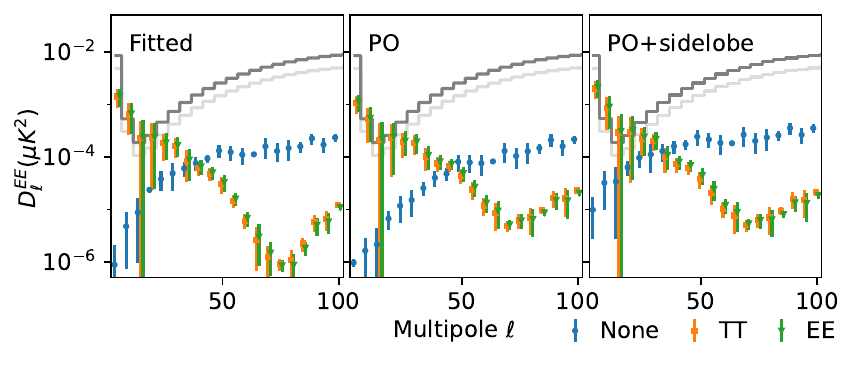}
\caption{
$EE$ power spectrum of residual dust maps due to beam mismatch. 
The light grey line indicates the binned sample variance for $f_{\textrm{sky}}=0.7$, the dark grey line for $f_{\textrm{sky}}=0.4$. 
Each sub-figure corresponds to a different beam model, indicated at the top. Within a sub-figure, each color and marker style refers to different gain calibrations. No calibration is attempted for the blue round markers (we assume the HWP-related gain is 1),  while $TT$ and $EE$ refer to the self-calibration $G^{XY}$ of Section~\ref{ssec:gaincal}. }
\label{fig:dust_beam}
\end{figure}
When subtracting maps simulated with the Gaussian beam from maps simulated with the fitted beam model or the PO beams, this results in some residuals in the $EE$ power spectrum.
We can observe those in Figure~\ref{fig:dust_beam}, where we plot the power spectrum of the residual dust maps up to $\ell=100$.
If we do not calibrate the HWP-related gain (i.e. $G=1$), we see residuals that increase fairly monotonically until $\ell\sim 50$ and then plateau, more marked for the PO+sidelobe case than the fitted or PO case.
This is consistent with residuals from the difference maps getting deconvolved by the $B_\ell$ associated with the main lobe (which is a more inadequate approximation in the PO+sidelobe case).
In the calibrated cases, we can see the bias scenario we described in Section~\ref{ssec:gaincal}: residuals for both $TT$ and $EE$ calibration are suppressed between $\ell=50$ and $\ell=100$ compared to the un-calibrated case, but are enhanced at low multipoles.
The value in the $7\leq \ell < 12$ bin can be seen in Table~\ref{tab:beam_syst_table}. 
The calibration mismatch leads to similar residuals for the $TT$ and $EE$ calibrations, which is to be expected since the HWP is kept ideal.
For CMB simulations, the residuals are about a percent of the sample variance limit for the three beam models. 
We do not see a difference between $TT$ and $EE$ based calibration, as the ideal behavior of the HWP stops any gain difference between those two spectra.

\begin{table}
\centering
\begin{tabular}{|c|c|c|c|}
\hline Sky, calibration & Fitted Beam & PO Beam & PO+Side\\\hline
CMB, no cal. &$<\!0.001$ &$<\!0.001$ & 0.001\\\hline
Dust, no cal. & 0.007 & 0.002 & 0.049\\\hline
CMB, $TT$ cal. & 0.019 & 0.015 & 0.025\\\hline
Dust, $TT$ cal. & \textbf{1.021} & 0.810 & \textbf{1.318}\\\hline
CMB, $TT$ cal. & 0.018 & 0.015 & 0.025\\\hline
Dust, $EE$ cal. & \textbf{1.051} & 0.854 & \textbf{1.399}\\\hline
\end{tabular}
\caption{Ratio of $EE$ power spectrum residuals to sample variance in the $7\leq \ell <12$ bin, as a fraction of sample variance for $f_\textrm{sky}=0.44$, due to substraction of a simulation with Gaussian beams. We bold the configurations where the residuals are larger than the sample variance.}
\label{tab:beam_syst_table}
\end{table}

For the dust map, the residuals are of the order of our sample variance target. 
On top of that, we see larger residuals for cases of the fitted beam and the physical optics beam with sidelobes, highlighting that sidelobes beyond a few FWHMs will significantly impact component characterization and separation.
This stresses the importance of beam characterisation and calibration for CMB experiments like Taurus.

\subsection{Half-wave plate non-idealities}\label{ssec:hwp_nonid}
We evaluate residuals due to HWP non-idealities for each of the beam models. 
As we are now probing a frequency-dependent effect, we apply the band-averaging method described in Section~\ref{ssec:beamconv}.

In Table~\ref{tab:hwp_residuals} we see that, if the input beam is fully known, the performance of the HWP is only weakly dependent on the specific beam model chosen: for a given HWP, sky component, and gain determination method, the residuals are similar for all four beam models. 
Beyond that, we are able to see clear differences in performance between the three HWP models depending on the choice of calibration scheme. 
The residuals for the single layer model, in the top two rows, are about six to seven times as large as for the three- and five-layer AHWPs in the absence of gain calibration. 
When calibrating the gain on $TT$, the residuals decrease for the three HWP models: moderately for BR1 and fairly aggressively for BR3 and BR5.
Calibrating on $EE$, the smallest residuals for the CMB simulation are achieved by the BR3 model, while the BR1 and BR5 models have percent-level residuals for the dust simulations.
This suggests the BR5 model is the most suitable if Taurus plans to calibrate on the $TT$ spectrum, and that the polarisation efficiency of both achromatic models can be calibrated effectively against the $EE$ power spectrum.

\begin{table}
\begin{center}
\resizebox{1 \textwidth}{!}{
\begin{tabular}{|c||c|c|c||c|c|c||c|c|c||c|c|c|}
\hline
    & \multicolumn{3}{c||}{Gaussian Beam} & \multicolumn{3}{c||}{Fitted Beam} & \multicolumn{3}{c||}{PO Beam} & \multicolumn{3}{c|}{PO+Sidelobe} \\
\hline
HWP, Sky & No & $TT$ & $EE$ & No & $TT$ & $EE$ & No & $TT$ & $EE$ & No & $TT$ & $EE$ \\\hline
BR1, CMB & 0.198 & 0.148 & 0.008 & 0.197 & 0.147 & 0.008 & 0.200 & 0.149 & 0.007 & 0.198 & 0.148 & 0.007\\\hline
BR1, Dust & \textbf{9.645} & \textbf{7.015} & 0.014 & \textbf{9.604} & \textbf{6.986} & 0.014 & \textbf{9.940} & \textbf{6.489} & 0.026 & \textbf{9.420} & \textbf{7.017} & 0.014\\\hline
BR3, CMB & 0.028 & 0.003 & 0.001 & 0.027 & 0.003 & 0.001 & 0.028 & 0.003 & 0.001 & 0.028 & 0.003 & 0.001\\\hline
BR3, Dust & \textbf{1.761} & 0.224 & 0.041 & \textbf{1.753} & 0.223 & 0.041 & \textbf{1.738} & 0.208 & 0.043 & \textbf{1.656} & 0.253 & 0.045\\\hline
BR5, CMB & 0.027 & 0.002 & 0.002 & 0.027 & 0.002 & 0.002 & 0.028 & 0.002 & 0.002 & 0.028 & 0.002 & 0.002\\\hline
BR5, Dust & \textbf{1.560} & 0.009 & 0.008 & \textbf{1.554} & 0.009 & 0.008 & \textbf{1.495} & 0.009 & 0.008 & \textbf{1.452} & 0.013 & 0.010\\\hline
\end{tabular}
}
\caption{Residuals due to HWP non-idealities for all four beam configurations. Each set of three columns is a different beam configuration, each line corresponds to a HWP and sky model. “No”, “$TT$” and “$EE$” refer to different estimation methods of the gain as defined in Section~\ref{ssec:gaincal}. Results are expressed as a fraction of the sample variance in the $7\leq \ell <12$ bin. We bold the configurations where the residuals are larger than the sample variance.}
\label{tab:hwp_residuals}
\end{center}
\end{table}

\subsubsection{Half-wave plate angle error}
\label{sec:hwpangerr}
The frequency-dependent phase shift of the AHWP models leads us to question another HWP-related non-ideality: what happens if the angle determination of the HWP is wrong? 
For an ideal HWP, an extra rotation of the HWP by an angle $\alpha$ is degenerate with a rotation of the polarisation-sensitive detectors by an angle $2\alpha$ \cite{Salatino2011}. 
In a real HWP with a non-diagonal Mueller matrix $M$, the $M_{IQ}, \, M_{IU}, \, M_{QI}$ and $M_{UI}$ terms will also be modulated by the rotation of the HWP, therefore an error on the HWP angle will propagate differently than an error on the detector polarisation angle. 
We evaluate the effect of a \SI{0.5}{\degree} error in determining the angle of each HWP model for all four beam models.
The CMB results, visible in Table~\ref{tab:hwp_angle_offset}, are nearly identical to those achieved with a detector polarisation angle offset in Section~\ref{ssec:point}, suggesting that the impact of the HWP angle error on the modulation of the off-diagonal terms is minimal.
However, we can observe that a half-degree error on the HWP's angle determination leads to residuals in the dust power spectrum that are more than half of the sample variance. 
This once again stresses the importance of calibration, this time of the HWP's angles.
We note however that the error we chose is a very conservative number considering the performance achieved by SPIDER \cite{Bryan2016} where the HWP rotation mechanism's angle was known to $\pm\SI{0.1}{\degree}$.

\begin{table}
\begin{center}
\resizebox{1 \textwidth}{!}{
\begin{tabular}{|c||c|c|c||c|c|c||c|c|c||c|c|c|}
\hline
    & \multicolumn{3}{c||}{Gaussian Beam} & \multicolumn{3}{c||}{Fitted Beam} & \multicolumn{3}{c||}{PO Beam} & \multicolumn{3}{c|}{PO+Sidelobe} \\\hline
HWP, Sky & No & $TT$ & $EE$ & No & $TT$ & $EE$ & No & $TT$ & $EE$ & No & $TT$ & $EE$\\\hline
BR1, CMB & -0.004 & -0.004 & -0.004 & -0.004 & -0.004 & -0.004 & -0.004 & -0.004 & -0.004 & -0.004 & -0.004 & -0.004\\\hline
BR1, Dust & 0.501 & 0.501 & 0.502 & 0.499 & 0.499 & 0.500 & 0.513 & 0.513 & 0.516 & 0.513 & 0.514 & 0.517\\\hline
BR3, CMB & -0.005 & -0.005 & -0.005 & -0.005 & -0.005 & -0.005 & -0.005 & -0.003 & -0.003 & -0.005 & -0.005 & -0.005\\\hline
BR3, Dust & 0.573 & 0.573 & 0.574 & 0.570 & 0.570 & 0.572 & 0.647 & 0.681 & 0.700 & 0.588 & 0.591 & 0.595\\\hline
BR5, CMB & -0.005 & -0.005 & -0.005 & -0.005 & -0.005 & -0.005 & -0.005 & -0.005 & -0.005 & -0.005 & -0.005 & -0.005\\\hline
BR5, Dust & 0.574 & 0.574 & 0.575 & 0.572 & 0.572 & 0.573 & 0.619 & 0.634 & 0.646 & 0.592 & 0.594 & 0.599\\\hline
\end{tabular}
}
\caption{Residuals in the $7\leq \ell <12$ bin, as a fraction of sample variance, due to \SI{.5}{\degree} HWP angle error (see Section \ref{sec:hwpangerr}). Lines and columns follow the definitions from Table~\ref{tab:hwp_residuals}. The small negative values are due to cut-sky effects within Namaster and are consistent with zero power.} \label{tab:hwp_angle_offset}
\end{center}
\end{table}

\subsection{Beam and half-wave plate interplay}
\label{sec:beamnhwmp}
What if we have imperfect knowledge of both the beam and the HWP model?
In Table~\ref{tab:hwp_beam_residuals} we show the residuals versus simulations made with Gaussian beams and ideal HWPs.
In the absence of gain calibration, the residuals are similar to those of Table~\ref{tab:hwp_residuals}, suggesting that the HWP non-ideality is the main systematic.
That is sensible as the residuals for the uncalibrated cases in Table~\ref{tab:beam_syst_table} are minor compared to the residuals caused by the HWP.

\begin{table}
\centering
\resizebox{1 \textwidth}{!}{
\begin{tabular}{|c||c|c|c||c|c|c||c|c|c|}
\hline
 & \multicolumn{3}{c||}{Fitted Beam} & \multicolumn{3}{c||}{PO Beam} & \multicolumn{3}{c|}{PO+Sidelobe} \\\hline
HWP, Sky & No & $TT$ & $EE$ & No & $TT$ & $EE$ & No & $TT$ & $EE$\\\hline
BR1, CMB & 0.206 & 0.070 & 0.024 & 0.204 & 0.078 & 0.023 & 0.223 & 0.062 & 0.032\\\hline
BR1, Dust & \textbf{10.003} & \textbf{2.995} & \textbf{1.058} & \textbf{10.226} & \textbf{2.992} & \textbf{1.096} & \textbf{10.685} & \textbf{2.649} & \textbf{1.361}\\\hline
BR3, CMB & 0.031 & 0.009 & 0.020 & 0.029 & 0.007 & 0.018 & 0.038 & 0.014 & 0.028\\\hline
BR3, Dust & \textbf{1.919} & 0.355 & \textbf{1.036} & \textbf{1.854} & 0.269 & 0.889 & \textbf{2.198} & 0.478 & \textbf{1.287}\\\hline
BR5, CMB & 0.031 & 0.022 & 0.019 & 0.029 & 0.018 & 0.017 & 0.038 & 0.028 & 0.027\\\hline
BR5, Dust & \textbf{1.718} & 0.960 & \textbf{1.042} & \textbf{1.606} & 0.759 & 0.848 & \textbf{1.975} & \textbf{1.160} & \textbf{1.290}\\\hline
\end{tabular}
}
\caption{$7\leq \ell <12$ multipole bin of the $EE$ power spectrum, as a fraction of the sample variance, for residual maps of the HWP+beam systematic (see Section~\ref{sec:beamnhwmp}).}\label{tab:hwp_beam_residuals}
\end{table}

When performing gain calibration, the residuals associated with the BR1 plate are divided by a factor of 3 when calibrating on $TT$ and a factor of 10 when calibrating on $EE$, making the residuals for the dust map equivalent in size to the sample variance associated with $f_\textrm{sky}=0.44$. 
Interestingly, this makes the $TT$ residuals smaller than in the case where we evaluate only the effect of HWP non-ideality.

For the BR3 HWP, the residuals are lowest when calibrating on the $TT$ power spectrum, no matter the beam model chosen. 
In fact, it is the only HWP and calibration configuration with residuals smaller than the cosmic variance target for the “PO+Sidelobe” beam model.
The “Fitted” and “PO+Sidelobe” beam models create higher residuals than in the case of a PO beam with no sidelobes, suggesting that characterizing sidelobes should be a more important focus for Taurus than getting an exact mapping of the main beam. 
The $\sim\SI{30}{\percent}$ difference between the residuals of the PO+sidelobes model and the azimuthal-symmetric fitted beam do however concede that there is some importance to non-azimuthally symmetric beam modes, and that a calibration campaign should not ignore them.

For the BR5 HWP model, gain calibration has trouble reducing the power spectrum residuals due to the beam mismatch issue.
Whereas, in Table~\ref{tab:hwp_residuals}, $TT$ or $EE$ calibration reduced the dust map residuals to less than \SI{1}{\percent} of the sample variance, here the residuals are between 1 and 1.3 times the sample variance depending on the beam model, with slightly smaller residuals when calibrating on $TT$.
The PO beam without sidelobes has once again smaller residuals, reinforcing the idea that it is the beam mismatch at low $\ell$ that makes calibration difficult.

\section{Conclusion}\label{sec:conclusion}

On average, the residuals we see at power spectrum level for the CMB sky scans are smaller than the target we set ourselves at the start of this study: that the systematic errors due to optical effects be often smaller than the cosmic variance associated with the \textit{Planck} $EE$ spectrum at a \SI{44}{\percent} sky fraction. 
In general, being able to reliably calibrate on the $EE$ power spectrum at smaller angular scales reduces residuals for HWP-related effects. 
The non-ideality with the largest effect, beam model mismatch, is unfortunately insensitive to that approach. 
As could be expected, replacing the single layer HWP with an AHWP increases the polarisation modulator's performance when scanning the CMB, although calibrating on the $EE$ power spectrum brings the behavior of the one-layer model in line with the two AHWPs. 

However, we should interrogate the possibility of such a calibration: it relies on having a simulation of Taurus' observations of the sky with a perfect instrument, but how will we know what the true sky looks like?
If we use \textit{Planck's} $EE$ spectrum, then we will be implicitly relying on \textit{Planck}'s model of its polarisation sensitivity for our analysis.
This problem is more important for the three-layer AHWP, as its scans had smaller residuals when calibrating on $EE$ than on $TT$.
The five-layer AHWP therefore seems like the best option, as it performs very well when calibrating on the temperature power spectrum, with the residual power in the $7\leq \ell \leq12$ bin of the $EE$ power spectrum at most \SI{1}{\percent} of the sample variance in that bin for dust foregrounds. 
This is an easier feat to accomplish for us: we could conceivably develop a reliable temperature calibration from the CMB dipole, or use \textit{Planck}'s $TT$ spectrum but retain freedom for our polarisation analysis.

The other concerning effect appears to be the coupling of sidelobes to the polarized dust emission of our galaxy. 
We note that imperfect knowledge of the beam model, and in particular of its far-sidelobes, is one of the main sources of systematic error for LiteBIRD's $B$-mode effort \cite{Litebird2023}. 
Taurus should have, at a minimum, an estimate of its average sidelobe amplitude within the instrument model akin to the fitted azimuthally symmetric beam we used in this paper. 
This is only made more urgent by the fact that $TT$ or $EE$ calibration increases the amplitude of the beam residuals, to the point where the residuals associated with the five-layer AHWP are larger than the sample variance limit.

On the other hand, we have only simulated the \SI{150}{\giga\hertz} band and are not forecasting how successful Taurus' component separation efforts will be once the four bands are included. 
We will however note that other collaborations that have either conducted simulations \cite{Shi2023, LSPE2021} or gathered data \cite{Li2023} appear to state with confidence that beam systematics can be kept under control when studying large-scale CMB polarisation by using astrophysical sources to characterize the beam and the polarisation angle. 
Holographic measurements \cite{Chesmore2022} could also provide pre-flight beam characterization.
\paragraph{General conclusion}
In summary, we have simulated a number of systematics related to the beams or the HWP for a one-month flight of the Taurus balloon. 
Using both Gaussian and PO beam simulations of 49 detector pairs in the \SI{150}{\giga\hertz} band, we generated timestreams and maps of simulated CMB and dust emission with \textit{beamconv} and evaluated the impact of non-idealities in the $EE$ power spectrum. 
We saw that the pointing and polarisation angle accuracy targets of Taurus led to negligible amounts of error in the $EE$ spectrum. 
We have found that the three HWP models perform differently, with the five-layer AHWP performing the best if the beam is fully known while the three-layer AHWP seems to deal with beam mismatch better.
We also observed that the coupling of beam sidelobes to polarized dust emission makes it difficult to self-calibrate the HWP polarisation efficiency if a Gaussian beam model is assumed. 
In future work, we would like to consider the impact of other systematics, like optical loading, and include more realistic timestreams with correlated detector noise.
We will also include the other frequency bands of Taurus. 

\acknowledgments

We would like to thank Nadia Dachlythra, Adri Duivenvoorden and David Alonso for productive discussions about \textit{beamconv} and Namaster. 
Some of the results in this paper have been derived using the \texttt{healpy} and HEALPix package.
Taurus is supported in the USA by NASA award number 80NSSC21K1957.
JEG acknowledges support from the Swedish Research Council (Reg.\ no.\ 2019-03959) and the Swedish National Space Agency (SNSA/Rymdstyrelsen). This work is in part funded by the European Union (ERC, CMBeam, 101040169). 


\bibliographystyle{JHEP}
\bibliography{biblio.bib}

\end{document}